# Size confinement effect in graphene grown on 6H-SiC (0001) substrate

V.M. Mikoushkin [1], V.V. Shnitov [1], A.A. Lebedev [1,2], S.P. Lebedev [1,2], S.Yu. Nikonov [1], O.Yu. Vilkov [3], T. Iakimov [4], R. Yakimova [4]

*V.Mikoushkin@mail.ioffe.ru*
[1] Ioffe Institute, 194021, St. Petersburg, Russia
[2] University ITMO, 197101, St. Petersburg, Russia
[3] Technische Universität Dresden, D-01062 Dresden, Germany
[4] Linkoping University, S-581 83, Linkoping, Sweden

We have observed the energy structure in the density of occupied states of graphene grown on n-type 6H-SiC (0001). The structure revealed with photoelectron spectroscopy is described by creation of the quantum well states whose number and the energy position ($E_1 = 0.3$ eV, $E_2 = 1.2$ eV, $E_3 = 2.6$ eV) coincide with the calculated ones for deep ($V = 2.9$ eV) and narrow ($d = 2.15$ Å) quantum well formed by potential relief of the valence bands in the structure graphene/n-SiC. We believe that the quantum well states should be formed also in graphene on dielectric and in suspended graphene.
PACS numbers: 73.20.At, 73.21.-b, 79.60.-i

After isolating graphene monolayer in 2004 [1], a lot of effort has been paid to study the fundamental properties of this unique object [2-5]. The most important of these properties is connected with confinement of the carrier motion, resulting in near liner dispersion for carrier energy and, as a consequence, in extremely small effective mass and huge carrier mobility along the 2D graphene plane. A question has arisen whether the 2D nature of graphene or few-layer graphene results in other size confinement peculiarities in the electronic structure, for example, those similar to the states in the comprehensively studied traditional semiconductor quantum wells. Indeed, van Hove singularities typical for semiconductor or metallic quantum wires were revealed in other low dimensional carbon system: in a single wall carbon nanotube which is a 1D system [6,7]. These singularities were found both in metallic and in semiconducting carbon nanotubes. Of particular interest is a possibility of size confinement effects in graphene or bilayer graphene grown on SiC substrate. The epitaxial graphene growth based on high temperature annealing of SiC substrate is considered to be one of the most promising technologies for wafer-scaled and high quality graphene film production [8,9]. In addition, charge transfer between graphene layer and substrate shifts Dirac point drastically increasing the carrier concentration and creating some bandgap, which is necessary for many applications [10-15]. At the same time, the system of multilayer graphene on n-type SiC wide bandgap semiconductor possesses a potential relief of a hole quantum well in which a structure of quantum well levels might be formed. Monolayer graphene on n-SiC also has a potential relief of a hole quantum well, though the width of the well seems too small to form even one energy level.

In this paper we report observation of an energy structure in the density of occupied states in single and bilayer graphene grown on n-type SiC (0001), which can be described by creation of the quantum well states. Formation of the states was shown to be possible due to large an electron/hole mass in the direction perpendicular to graphene plane.

The graphene film was grown on n-type 6H-SiC (0001) silicon carbide substrate by sublimation method with an original pre-growth treatment essentially enhancing the film quality [13,14,16]. The growth was performed on nominally on-axis 6H-SiC wafers with polished Si (0001) face purchased from Cree Corp. The high-vacuum annealing of SiC wafers before graphene growth was used for removing distorted surface layer caused by mechanical polishing [16]. The graphene film formations was carried out in inductively heated furnace at temperature of 2000°C and at an ambient argon pressure of 1 atm. Properties of epitaxial graphene were studied *ex situ* by atomic force microscopy (AFM), low energy electron diffraction (LEED), x-ray photoelectron spectroscopy (XPS), and near edge x-ray absorption fine structure (NEXAFS) spectroscopy. AFM study showed that substrate surface consists of flat and wide (~1 μm) terrac-es covered with sufficiently large and continuous graphene domains [Fig. 1(a)]. Numerous LEED patterns obtained from different points of the sample demonstrate concurrent presence of a well-ordered graphite (1×1) pattern and (6√3×6√3)R30 pattern inherent to the underlying buffer layer [10-12] and, thereby, evidencing formation of thin graphene layer consisting of one-two monolayers [Fig. 1(b)].

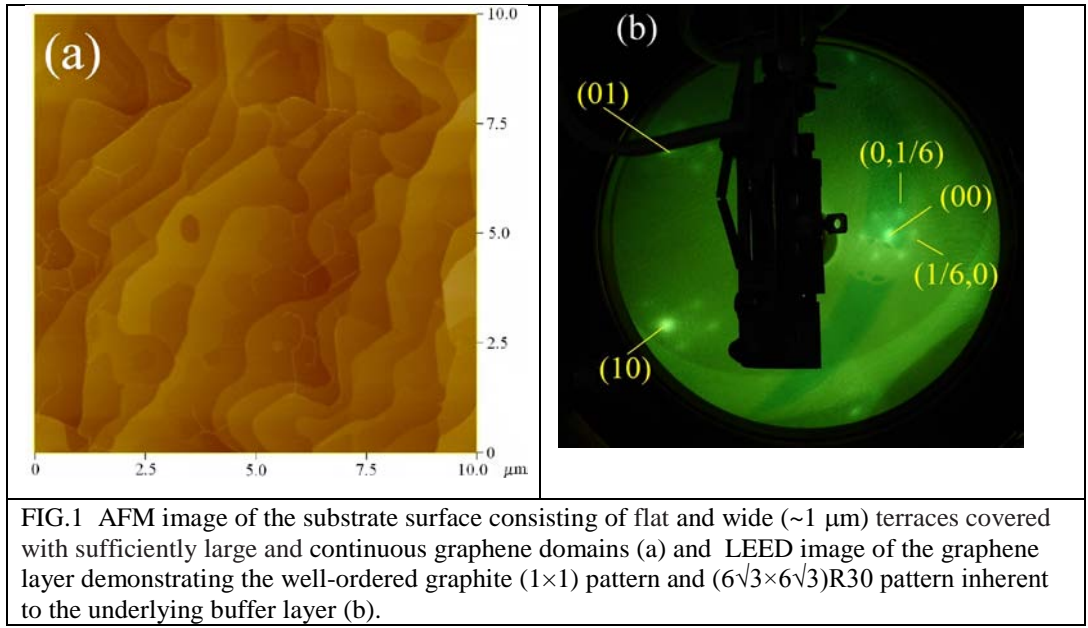

FIG.1 AFM image of the substrate surface consisting of flat and wide (~1 μm) terraces covered with sufficiently large and continuous graphene domains (a) and LEED image of the graphene layer demonstrating the well-ordered graphite (1×1) pattern and (6√3×6√3)R30 pattern inherent to the underlying buffer layer (b).

The chemical composition and electronic structure of the samples were studied by x-ray photoelectron spectroscopy (XPS) at the Russian–German synchrotron radiation beamline of the BESSY-II electron storage ring [17]. The photoelectron spectra were measured at normal direction of photoelectron emission using a SPEC hemispherical analyzer with a total spectral resolution of $\Delta E = 0.15$ eV. X-ray incidence angle was $60^o$. The energy scale of the photoelectron spectra was periodically calibrated using the Au $4f_{7/2}$ photoelectron line of a gold reference sample. The error of determining the binding energies for the core electron levels studied did not exceed $\delta E_b = 0.05$ eV. Prior to XPS measurements, the surface of samples was cleaned by prolonged (for many hours) keeping in vacuum followed by one hour low-temperature (T = 250°C) heating in high vacuum [18,19].

Accurate determination of the layer thickness is of particular importance in the analysis of the graphene physical properties. This task was solved in our research directly in the studied area of the sample by analyzing the line intensities of C1s core level XPS spectra. One of the spectra is shown in Fig. 2 together with the spectrum of highly oriented pyrolytic graphite (HOPG). The spectrum shown is similar to the C1s spectra obtained earlier for analogous graphene/SiC systems [10,12,13]. Line *G* corresponds to C=C bonds (or sp$^2$ bonding) in graphene. Photon energy ($h\nu$ = 450 eV) and corresponding photoelectron kinetic energies were high enough to detect the contributions of the buffer layer *S2* and even the substrate *SiC*. The line shift (~0.1 eV) to higher binding energies relatively to HOPG indicates the charge transfer between graphene and substrate, which results in the same shift of the Dirac point below the Fermi level. This shift proved to be a bit less in our research as compared to the earlier observations [11,12] for graphene films with one-two monolayer thickness typical also for this work (see below).

The film thickness (*d*) averaged throughout the studied spot (1*0.2 mm) was measured by comparing the relative C1s line intensities of graphene ($I_G$) and SiC substrate ($I_{SiC}$). The method used was expected to be rather accurate in our case of the carbon containing substrate (SiC) since all the data were obtained in one experiment with one sample in the narrow photoelectron energy range. Therefore neither information about spectrometer transmission nor data about surface roughness are needed. Relative C1s line intensities in graphene layer capping a semi-infinite SiC substrate can be given by the following relation:

$$\frac{I_G}{I_{SiC}} = \frac{\sigma'_G \, T_G \, n_G \, \lambda_G \, [1-\exp(-d/\lambda_G)]}{\sigma'_{SiC} \, T_{SiC} \, n_{SiC} \, \lambda_{SiC} \, \exp(-d/\lambda_G)} \quad , \quad (1)$$

where $\sigma' = d\sigma/d\Omega$ is the differential cross-section of C1s core level photoemission, $T$ is the transmission function of the analyzer, $n$ is the density of carbon atoms, and $\lambda$ is the inelastic mean free path. Indices *"G"* and *"SiC"* point out to belonging the terms to graphene layer and substrate, correspondingly. The C1s cross-sections $\sigma'_G$ and $\sigma'_{SiC}$ can be accepted equal, as well as the transmission functions $T_G$ and $T_{SiC}$ for photoelectrons with close energies. The quotient $(n_G *\lambda_G) / (n_{SiC} *\lambda_{SiC}) = 2.1$ for C1s photoelectron energy E =165 eV corresponding to photon energy $h\upsilon$ = 450.0 eV used in this work, taking into account the values $\lambda_G$ =5.4 Å [20,21] and $\lambda_{SiC}$ =5.8 Å [22] in graphite and SiC, correspondingly. Therefore the graphene layer thickness *d* can be found with sufficient accuracy by the formula:

$$d = \lambda_G \, ln(\frac{I_G}{2.1*I_{SiC}} + 1) \, . \quad (2)$$

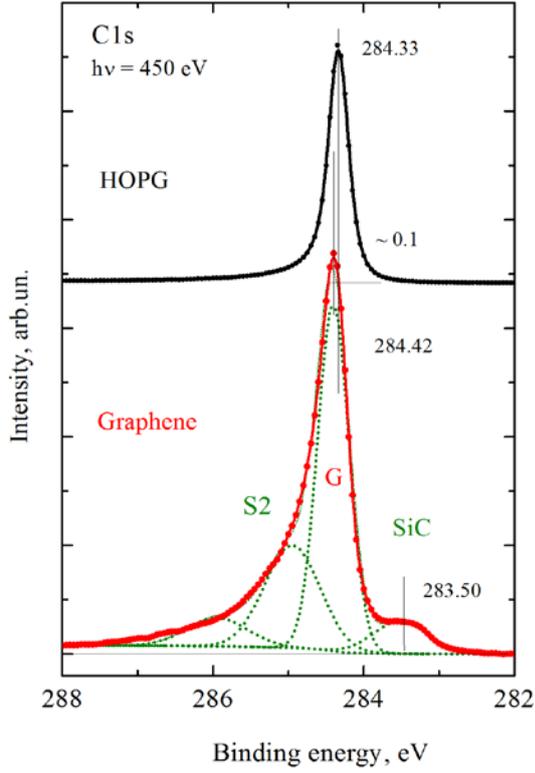

FIG. 2 C1s photoelectron spectra of HOPG and graphene on n-type 6H SiC (0001) substrate.

The thickness *d* of the graphene layer was obtained in experiments with photon energies $h\upsilon$ = 450.0 and 700 eV. The result and the data used are presented in Table 1. The interlayer spacing in graphene was taken to be equal to that in graphite (3.35 Å). Since the capping layer covering the substrate (SiC) contains the buffer layer (6√3*6√3)R30 underneath the graphene, the sum of the line intensities $(I_G+I_S)$ was used for the capping layer instead of the intensity $I_G$ in formula (2). Areas under the corresponding lines were used instead of the line intensities (peak hights) to avoid the error connected with different widths of different lines. The thickness of the buffer layer of 2.4 Å was taken from Ref. [12]. Table 1 shows graphene thicknesses in monolayers (ML) calculated both for mean free path $\lambda_G$ [20,21] and attenuation length $\lambda'_G$ [23]. The last one was obtained by correction of the first one with taking into account the attenuation due to elastic scattering [23]. The average thickness of the graphene layer grown in this research proved to be $d \approx 1.6$ ML. It means that manolayer and bilayer graphene occupy approximately equal areas. The error of the graphene thickness is mainly determined by that of $\lambda_G$ calculations [20,21,23], that is $\varepsilon(d) = \pm 20\%$.

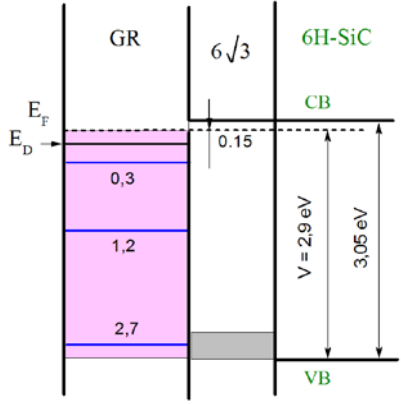

FIG. 3 Scheme of the hole quantum well formed by graphene on SiC with 6√3 interface carbon monolayer.

TABLE 1. Effective graphene layer thickness deduced from relative intensities of C1s lines for graphene layer $I_G$, substrate $I_{SiC}$, and interface layer $I_S$.

| hν, eV | $I_G$ | $I_{SiC}$ | $I_S$ | $\lambda_G$, Å  $d$, ML | $\lambda'_G$, Å  $d'$, ML |
|---|---|---|---|---|---|
| 450.0 | 6.1 | 1.0 | 3.12 | 5.4  **2.0** | 4.3  **1.4** |
| 700.0 | 2.1 | 1.0 | 1.4 | 9.6  **2.1** | 8.2  **1.7** |

The peculiarity of the system studied is that graphene layer is placed on the wide bandgap n-type semiconductor, creating a layered nanostructure with the hole quantum well (QW) potential relief. The scheme of such a structure is shown in Fig. 3. The buffer layer (6√3*6√3) R30 underneath the graphene layer is considered in this scheme also to be a wide bandgap layer since every third carbon atom of the layer is chemically bound with SiC substrate [12], which radically diminishes the π-electron subsystem and opens the bandgap. Top of the graphene valence band (shaded area) looks like a hole quantum well in this scheme. One can assume creation of the states (bands) in this quantum well due to size confinement of the electron/hole motion in normal direction (perpendicular to the graphene plane). These states should manifest itself as peaks in the valence band (VB) density of states (DOS) and VB photoemission spectra reflecting the DOS.

Fig.4 a shows valence band photoemission spectra of SiC substrate, graphene film on SiC substrate and phyrolytic graphite. Three peaks are seen in the graphene VB spectrum in the range near the Fermi level at energies $E_1 = 0.3$ eV, $E_2 = 1.2$ eV and $E_3 = 2.6$ eV, whereas no peculiarities are seen in the spectra of SiC substrate and graphite. We should notice that similar structure was observed in 1996 by Johansson, Owman, and Mårtensson [24] in studying the reconstruction of 6H-SiC (0001) surface, though it was assigned to (6√3* 6√3)R30° reconstruction. But the annealing temperature 1150°C was sufficient to create a graphene layer. Indeed, C1s core level spectra showed the "graphitic" component (284.7 eV) [24]. Therefore the structure considered should be assigned to graphene layer rather than to the underneath (6√3*6√3)R30° buffer layer. Moreover, there must be no states in the buffer layer near the Fermi layer at all because of the mentioned above radical depletion of the π-electron subsystem due to chemical binding of one third of the carbon atoms in the buffer layer with substrate [12], which should open the bandgap in the buffer layer. The residual photoelectron signal from the buffer layer is attenuated by 3-7

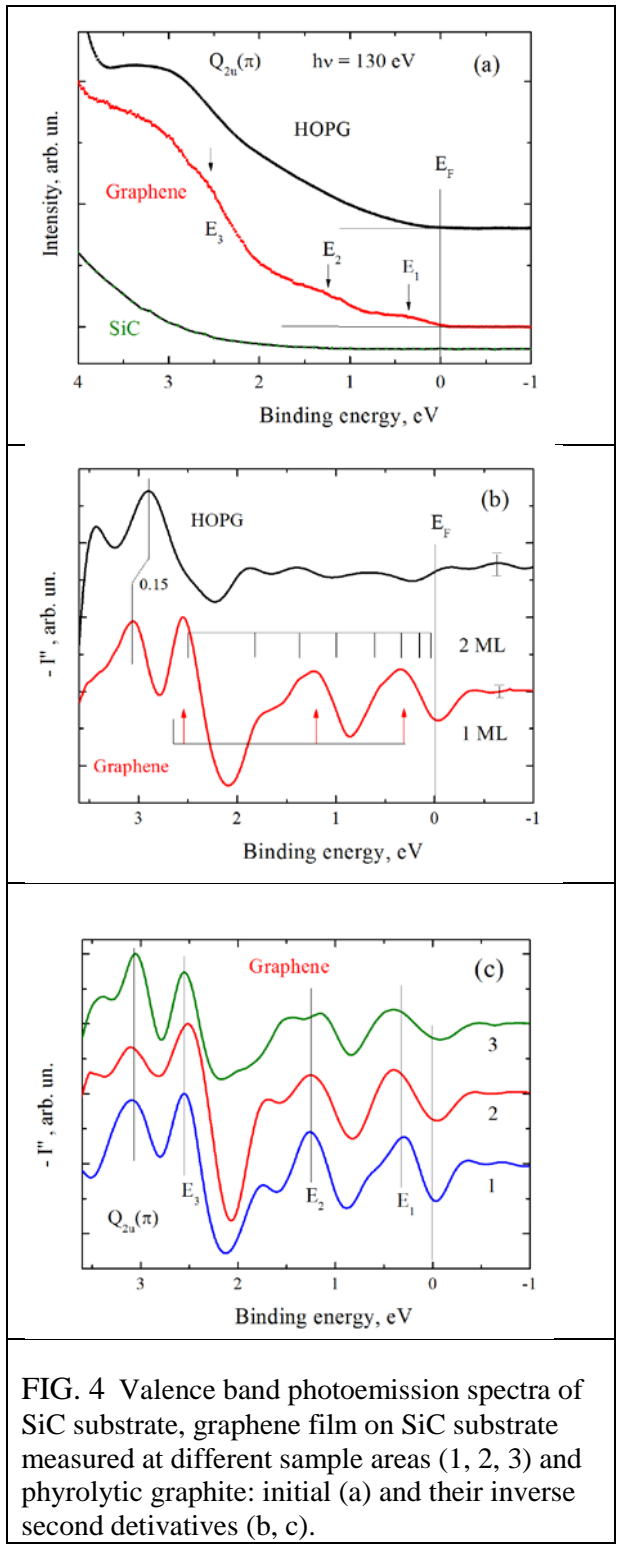

FIG. 4 Valence band photoemission spectra of SiC substrate, graphene film on SiC substrate measured at different sample areas (1, 2, 3) and phyrolytic graphite: initial (a) and their inverse second detivatives (b, c).

times in the graphene layer and does not contribute to the photoemission spectrum shown in Fig. 4.

To determine the energy position of the observed peaks more accurate, the VB spectra were differentiated. Fig. 4 b shows second derivative of the valence band photoemission spectrum of graphene film in comparison with that of graphite. The graphene spectrum is the sum of three spectra measured in different areas of the film. These spectra are shown in Fig. 4c. They illustrate high reproducibility of all the structure peculiarities. The summary spectrum in Fig. 4b shows the structure observed in each of the constituents, but with better statistics. Four peaks are seen in the first approximation. One of them (~ 3 eV) is related to the $Q_{2u}(\pi)$ state typical to gra-phite. This peak is shifted by almost the same value (~0.15 eV) as C1s photoemission line did due to charge transfer between graphene layer and substrate (Fig. 2).The energy positions of three other peaks coincide with the energies of three QW states calculated in approximation of asymmetric quantum well [25]:

$$E_n = \frac{\hbar^2 k_n^2}{2m}, \qquad (3)$$

$$\pi n - k_n d = \arcsin\left(\frac{\hbar k_n}{\sqrt{2mV}}\right), \qquad (4)$$

where $E_n$ is the QW energies, $k_n$ is the wave number, $d$ is QW width, $V$ is the barrier height at the interface and $m$ is the hole/electron effective mass.

The barrier height is $V = 2.9$ eV according to the scheme shown in Fig. 3. The effective mass was taken to be equal to the hole and electron masses in graphite $m_h = m_e = 22\, m_o$ for the normal direction [26]. Here $m_o$ is the electron mass. Taking these values, the following QW energies were obtained for monolayer graphene: $E_1 = 0.3$ eV, $E_2 = 1.2$ eV and $E_3 = 2.55$ eV. These values practically coincide with the positions of the most intensive three experimental peaks. The absolute agreement of calculated and experimental data was achieved at QW thickness $d = 2.15 \pm 0.05$ Å which is less than the interplane spacing in graphite (3.35 Å) usually used for estimation of one monolayer graphene thickness. But the one monolayer graphene thickness, indeed, should be less than the interplane graphite distance, since only one carbon monolayer sheet contributes to the density of the out-of-plane π-electrons.

Since the quantum well is very deep (2.9 eV), one could expect a satisfactory description of the QW structure in a simpler approximation of infinite symmetric quantum well:

$$E_n = \frac{\hbar^2}{2m}\left(\frac{\pi n}{d}\right)^2. \qquad (5)$$

Indeed, the QW energies calculated in this approximation proved to be practically the same at $d = 2.4$ Å : $E_1 = 0.3$ eV, $E_2 = 1.2$ eV and $E_3 = 2.7$ eV. Therefore this approximation was used to estimate the QW spectrum of bilayer graphene covered approximately a half of the sample area as was shown above. Positions of the QW levels obtained for bilayer graphene are shown in Fig. 4b. The QW width $d = 6.7$ Å for bilayer graphene occurred to be very close to the value estimated from interplane spacing in graphite (3.35*2 = 7 Å). The major part of the bilayer graphene states contributes to the main three peaks, but two of them ($E_4 = 0.6$ eV and $E_7 = 1.85$ eV) describe statistically reliable spectra peculiarities (peak shouldes). This fact confirms presence of bilayer graphene in studied film besides monolayer graphene. Contributions of QW spectra from monolayer and bilayer graphene satisfactory describe the structure observed in VB photoemission spectra. This fact evidences creation of the valence electron states in graphene due to confinement of their motion.

In summary, we revealed a new graphene property related to its 2D atomic structure: the ability to create a quantum well and quantum well levels. This ability was studied in mono- and bilayer graphene grown on n-type 6H-SiC (0001) wide bandgap semiconductor. We showed that the structure in the valence band density of states near the Fermi level is described by the quantum well states whose number and the energy position coincide with the calculated ones. We can assume that the revealed property is an attribute not only of graphene (or few-layer graphene) on the wide-bandgap semiconductor substrate but also that of graphene on dielectric and of suspended graphene. We can also conclude that the type of the semiconductor substrate should define the type of quantum well: n-type results to QW in occupied VB and p-type - to QW in unoccupied VB. The QW state formation becomes possible in such a narrow quantum well due to large electron/hole mass in the direction perpendicular to graphene plane.

The research was partly supported by the Russian-German Laboratory at BESSY II, by the FASR contract 02.740.11.0108 by the program "Quantum physics of condensed matter" of the Russian Academy of Sciences and RFBR project № 12-02-00165a. VR (Sweeden) is greatly acknowledged by RV and TI.